\documentclass{whitepaper}
\pdfoutput=1

\input{common/preamble}

\begin{document}

\pagestyle{titlepage}


\pagestyle{titlepage}

\date{}

\title{\scshape\Large Broader Impact of Cyclotron-based Neutrino Sources\\
\vspace{10mm}
\normalsize White Paper contribution to NF07\\ 
as part of the Snowmass'21 Community Planning Exercise
\vskip -10pt

}


\renewcommand\Authfont{\scshape\small}
\renewcommand\Affilfont{\itshape\footnotesize}

\author{Jose Alonso, for the IsoDAR Collaboration}
\date{\today}

\vspace{-0.5cm}

\maketitle

\renewcommand{\familydefault}{\sfdefault}
\renewcommand{\thepage}{\roman{page}}
\setcounter{page}{0}

\pagestyle{plain} 





\renewcommand{\thepage}{\arabic{page}}
\setcounter{page}{1}

\pagestyle{fancy}

\fancyhead{}
\fancyhead[RO]{\textsf{\footnotesize \thepage}}
\fancyhead[LO]{\textsf{\footnotesize \rightmark}}

\fancyfoot{}
\fancyfoot[RO]{\textsf{\footnotesize Snowmass'21}}
\fancyfoot[LO]{\textsf{\footnotesize NF07 White Paper contribution}}
\fancypagestyle{plain}{}

\renewcommand{\headrule}{\vspace{-4mm}\color[gray]{0.5}{\rule{\headwidth}{0.5pt}}}



\clearpage

\begin{abstract}
    The cyclotron designed for the IsoDAR neutrino source represents a paradigm shift in cyclotron performance -- a factor of 10 increase in beam current.  This performance is required to develop a neutrino source of sufficient strength for meaningful ``decay-at-rest'' experiments, but it also can have important ramifications for the isotope-production sector for medical or other applications, by enabling efficient, high-yield production of long-lived isotopes, or isotopes where production cross sections are low.  
\end{abstract}

\section{Introduction - Description of IsoDAR}


Much of the material presented here appears in other White Papers~\cite{AF02} and prior publications related to IsoDAR~\cite{Isophysics,Alonso:2022mup}. Here we only wish to highlight the societal benefits inherent in the significant developments that have enabled the large increases in beam current in what we believe will be a new generation of cyclotrons.

Cyclotrons can play a very important role as neutrino sources, by enabling nuclear reactions that produce unstable, beta-decaying isotopes.  Such isotopes, whether from reactors, sun, or monoisotopic sources, form a large fraction of the neutrino research sector.  Monoisotopic sources offer the advantage of a clean, well understood neutrino spectrum, but obtaining sources of suitable lifetimes that can be produced and transported to a large detector is not straightforward.  However, a very effective source can be obtained by continuously producing an isotope by means of a cyclotron beam and a target close to the detector.  

The IsoDAR experiment~\cite{Isophysics} is a noteworthy example.  In this experiment, to be deployed at the new Yemilab underground facility in Korea~\cite{Alonso:2022mup}, a 10 mA proton beam strikes a Be+D$_2$O target producing neutrons that flood into a meter-diameter sleeve containing a mixture of Be powder and highly enriched $^7$Li.  The decay of 839-millisecond $^8$Li, following neutron capture, provides the highly-favorable neutrino spectrum that is studied by the very close-by 2.3 kiloton liquid scintillator detector~\cite{Seo:LSC}.  The 10 mA of beam produces, effectively, a 50 kilocurie source of $^8$Li. Because of the short lifetime the source strength is closely coupled to the cyclotron beam current.  This current was specified by the need for statistics in this “decay-at-rest” configuration, where even placing the target as close as allowed to emplace adequate shielding to keep neutrons out of the detector (about 7 meters, filled almost entirely with steel and concrete), the detector only subtends about 5\% of the solid angle of the source.

\section{Impact of High Currents for Isotope Production}

This 10 mA represents a significant increase in the state-of-the-art current, a factor of 4 over the current record-holder, the 72 MeV Injector 2 booster separated-sector cyclotron at the Paul Scherrer Institute~\cite{PSI:Inj2}, and a factor of 10 over commercial isotope-producing cyclotrons.  Table~\ref{cyclodesigntable} compares the IsoDAR cyclotron with two commercial cyclotrons dedicated to isotope production.

\begin{table}[b]
\begin{center}
\caption{\footnotesize Comparison of IsoDAR with IBA commercial isotope cyclotrons.\label{cyclodesigntable}}
\renewcommand{\arraystretch}{1.25}
{
\begin{tabular}{|lccc|}
\hline
Parameter & IsoDAR & IBA C-30 & IBA C-70\\ 
\hline 
Ion species accelerated & H$_2^+$ & H$^-$ & H$^-$ \\
Maximum energy (MeV/amu) & 60 & 30 & 70 \\
Proton beam current (milliamps) & 10 & 1.2 & 0.75 \\
Available beam power (kW) & 600 & 36 & 52 \\
Pole radius (meters) & 1.99  & 0.91 & 1.24  \\
Outer diameter (meters) & 6.2 & 3 & 4 \\
Iron weight (tons) & 450 	& 50 & 140 \\
Electric power reqd. (megawatts) & 2.7 & 0.15 & 0.5 \\
\hline
\end{tabular}}
\end{center}
\end{table}

\begin{figure}[h]
\centering
\includegraphics[width=5in]{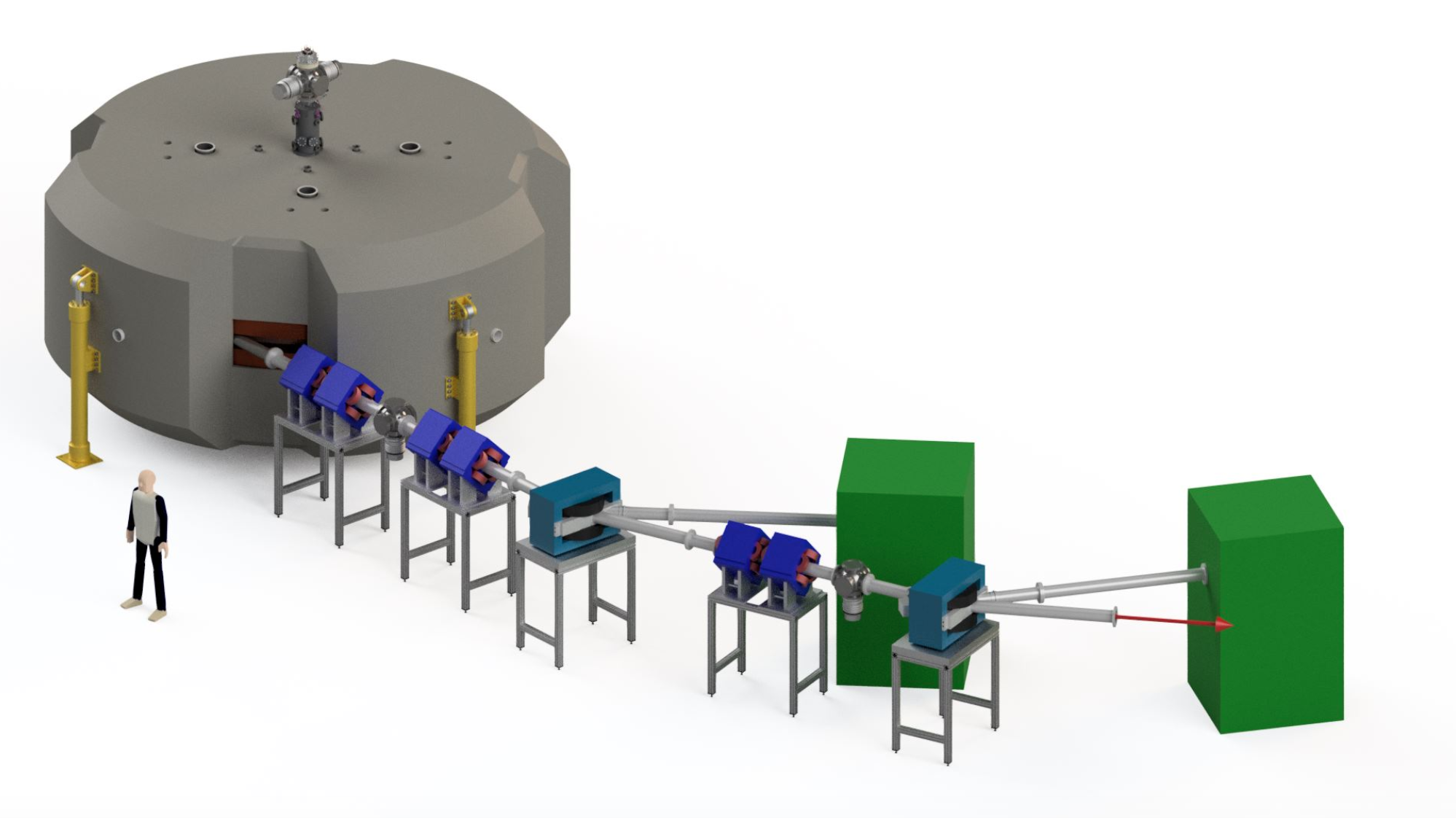}
\caption{{\footnotesize IsoDAR 60 MeV/amu cyclotron.  The compact size makes deployment for underground neutrino experiments quite efficient.  The figure shows a scheme for sharing high power beam with several target stations for isotope production, by pushing a stripper foil into the edge of the extracted H$_2^+$ beam just upstream of an analyzing dipole, shaving off the resulting protons into the target station.  Note the ion source, and RFQ buncher buried in the iron yoke at the center of the cyclotron. }\label{cyclotron}}
\vspace{0.2in}
\end{figure}

Such a current increase offers significant benefits for isotope production: enabling substantial increases in the production rate and source-strength for long-lived isotopes, or those produced in low-cross section reactions.  Detailed evaluations can be found in references~\cite{alonso:isotopes,waites:isotopes}.  Fully utilizing the very high beam powers will require development of suitable targets, and the ability to split beam amongst several target stations (see Fig.~\ref{cyclotron});  but, efficient utilization of the beam can have a huge impact on the availability of important isotopes, such as $^{68}$Ge~\cite{Ge/Ga} and $^{225}$Ac~\cite{Ac225}, that are at present difficult to produce, and are very expensive.   $^{68}$Ge is the 271 day parent of the $^{68}$Ge/$^{68}$Ga generator, the Ge is produced and shipped close to the use site at a hospital or imaging center.  The $^{68}$Ga (57 minute) daughter is ``milked'' from the Ge cell and used for PET imaging in the patient.  The long half-life of the parent makes it difficult to produce in the ~20 millicurie amounts needed to make a generator of sufficient intensity for good diagnostic applications. The long half life also increases the time the generator can produce useful doses of the PET isotope. Full beam utilization of the IsoDAR Cyclotron produces 1 curie of $^{68}$Ge in 3 hours  of cyclotron time (about 20 times the rate of an IBA C-30). $^{225}$Ac is a 10-day alpha emitter, feeding into a chain of decays, leading ultimately to stable $^{205}$Tl.  This chain includes four isotopes with alpha decays.  These four alphas, all emitted into the same cell where the radioactive nucleus is located form a lethal dose of short-ranged radiation.  It has proven clinically extremely effective against metastatic tumors~\cite{Ac225}, dependent on the carrier pharmaceutical to transport the activity to the tumor locations.  With our cyclotron, 1 curie of $^{225}$Ac can be produced in 5 hours of cyclotron time.  A dose is typically a few 10's of millicuries, so even after chemical processing to remove unwanted spallation-produced activities (most notably $^{227}$Ac~\cite{Robertson}), significant quantities of this isotope can be produced, far above other production avenues.

An additional feature of accelerating H$_2^+$, which has a charge-to-mass ratio of 1/2, is that this cycloton will accelerate other ions with the same 1/2 Q/A.  This includes fully-stripped helium nuclei ($\alpha$ particles), and C$^{6+}$.  These ions open up a large field of new isotopes for possible commercial application, and ion sources are commercially available with good intensities of these difficult-to-produce ions~\cite{Pantechnik}.  

The performance improvements for the cyclotron have arisen from several important developments, including  acceleration of H$_2^+$ ions~\cite{calabretta:htp,winklehner:mist1}, efficient pre-bunching of the beam~\cite{Winklehner:rfq},  utilization of the ``vortex’’ motion effect discovered at PSI~\cite{baumgarten:vortex1} whereby space-charge forces contribute to stabilization of the beam bunches in a cyclotron~\cite{Winklehner:2021qzp}.  The extensive modeling with sophisticated simulation codes and initial experimental work detailed in the above references have demonstrated that the performance levels for the IsoDAR cyclotron (proton beam on target of 10 mA at 60 MeV) are achievable with relatively low risk.  In addition, the key elements of the system: high-current ion source, RFQ buncher, capture and stable vortex formation all occur at the central region of the cyclotron, forming a ``Front End” system that can form the basis for a family of cyclotrons, with extracted beam energy tailored for specific applications.  Designs for energies between 15 and 60 MeV are straightforward.
Extending to energies beyond 60~MeV/amu is possible, but will need to be studied more carefully as the turn separation becomes smaller as energy increases.

\section{Conclusion}

In summary, isotope production via the IsoDAR cyclotron is an example of how technical developments required for successful neutrino experiments can have a broad societal impact.

\textbf{Acknowledgments}
\label{sec:acknowledgements}

Development of IsoDAR is supported by the National Science Foundation (NSF) and the Heising Simons Foundation.




\renewcommand{\refname}{References}

\printglossary

\bibliographystyle{utphys}


\bibliography{common/citedb}

\providecommand{\href}[2]{#2}\begingroup\raggedright\begin{thebibliography}{10}

\bibitem{AF02}
D.~Winklehner {\em et~al.} in {\em {Proceedings of the Snowmass Community
  Planning Exercise (Snowmass’2021)}}.
\newblock 2022.
\newblock \url{https://snowmass21.org/submissions/start}.

\bibitem{Isophysics}
J.~Alonso, C.~A. Arg\"uelles, J.~M. Conrad, Y.~D. Kim, D.~Mishins, S.~H. Seo,
  M.~Shaevitz, J.~Spitz, and D.~Winklehner
  \href{http://arxiv.org/abs/2111.09480}{{\ttfamily arXiv:2111.09480
  [hep-ex]}}.

\bibitem{Alonso:2022mup}
J.~R. Alonso {\em et~al.} \href{http://arxiv.org/abs/2201.10040}{{\ttfamily
  arXiv:2201.10040 [physics.ins-det]}}.

\bibitem{Seo:LSC}
S.-H. Seo \href{http://arxiv.org/abs/1903.05368}{{\ttfamily arXiv:1903.05368
  [physics.ins-det]}}.

\bibitem{PSI:Inj2}
A.~Kolano, A.~Adelmann, R.~Barlow, and C.~Baumgarten
  \href{http://dx.doi.org/https://doi.org/10.1016/j.nima.2017.12.045}{{\em
  Nuclear Instruments and Methods in Physics Research Section A: Accelerators,
  Spectrometers, Detectors and Associated Equipment} {\bfseries 885} (2018)
  54--59}.
  \url{https://www.sciencedirect.com/science/article/pii/S0168900217314511}.

\bibitem{alonso:isotopes}
J.~R. Alonso, R.~Barlow, J.~M. Conrad, and L.~H. Waites
  \href{http://dx.doi.org/10.1038/s42254-019-0095-6}{{\em Nature Reviews
  Physics} {\bfseries 1} no.~9, (Sept., 2019) 533--535}.
  \url{https://www.nature.com/articles/s42254-019-0095-6}.

\bibitem{waites:isotopes}
L.~H. Waites, J.~R. Alonso, R.~Barlow, and J.~M. Conrad
  \href{http://dx.doi.org/10.1186/s41181-020-0090-3}{{\em EJNMMI Radiopharmacy
  and Chemistry} {\bfseries 5} no.~1, (Feb., 2020) 6}.
  \url{https://doi.org/10.1186/s41181-020-0090-3}.

\bibitem{Ge/Ga}
F.~Rösch
  \href{http://dx.doi.org/https://doi.org/10.1016/j.apradiso.2012.10.012}{{\em
  Applied Radiation and Isotopes} {\bfseries 76} (2013) 24--30}.
  \url{https://www.sciencedirect.com/science/article/pii/S0969804312005350}.
  Ga-68 Special Issue.

\bibitem{Ac225}
C.~Kratochwil, F.~Bruchertseifer, F.~L. Giesel, M.~Weis, F.~A. Verburg,
  F.~Mottaghy, K.~Kopka, C.~Apostolidis, U.~Haberkorn, and A.~Morgenstern
  \href{http://dx.doi.org/10.2967/jnumed.116.178673}{{\em Journal of Nuclear
  Medicine} {\bfseries 57} no.~12, (2016) 1941--1944},
  \href{http://arxiv.org/abs/https://jnm.snmjournals.org/content/57/12/1941.full.pdf}{{\ttfamily
  https://jnm.snmjournals.org/content/57/12/1941.full.pdf}}.
  \url{https://jnm.snmjournals.org/content/57/12/1941}.

\bibitem{Robertson}
A.~K.~H. Robertson, B.~L. McNeil, H.~Yang, D.~Gendron, R.~Perron, V.~Radchenko,
  S.~Zeisler, P.~Causey, and P.~Schaffer
  \href{http://dx.doi.org/10.1021/acs.inorgchem.0c01081}{{\em Inorganic
  Chemistry} {\bfseries 59} no.~17, (2020) 12156--12165},
  \href{http://arxiv.org/abs/https://doi.org/10.1021/acs.inorgchem.0c01081}{{\ttfamily
  https://doi.org/10.1021/acs.inorgchem.0c01081}}.
  \url{https://doi.org/10.1021/acs.inorgchem.0c01081}. PMID: 32677829.

\bibitem{Pantechnik}
A.~Villari {\em et~al.} in {\em Proceedings of the DAE-BRNS Indian particle
  accelerator conference}.
\newblock 2011.

\bibitem{calabretta:htp}
L.~Calabretta, D.~Rifuggiato, and V.~Shchepounov in {\em {Proceedings of the
  1999 Particle Accelerator Conference: New York}}, p.~3288.
\newblock 1999.
\newblock \url{https://accelconf.web.cern.ch/p99/PAPERS/THP139.PDF}.

\bibitem{winklehner:mist1}
D.~Winklehner, J.~M. Conrad, J.~Smolsky, and L.~H. Waites.
  \url{https://aip.scitation.org/doi/10.1063/5.0063301}. Publisher: American
  Institute of Physics.

\bibitem{Winklehner:rfq}
D.~Winklehner \href{http://dx.doi.org/10.1016/j.nima.2018.07.036}{{\em Nucl.
  Instrum. Meth. A} {\bfseries 907} (2018) 231--243},
  \href{http://arxiv.org/abs/1807.03759}{{\ttfamily arXiv:1807.03759
  [physics.acc-ph]}}.

\bibitem{baumgarten:vortex1}
C.~Baumgarten {\em Physical Review Special Topics-Accelerators and Beams}
  {\bfseries 14} no.~11, (2011) 114201.

\bibitem{Winklehner:2021qzp}
D.~Winklehner, J.~M. Conrad, D.~Schoen, M.~Yampolskaya, A.~Adelmann, S.~Mayani,
  and S.~Muralikrishnan. \url{https://doi.org/10.1088/1367-2630/ac5001}.
  Publisher: {IOP} Publishing.

\end{thebibliography}\endgroup


\end{document}